\documentclass[aps,prl,showpacs]{revtex4}
\usepackage{graphicx}
\usepackage{mathrsfs}
\usepackage{amssymb}

\newcommand{\D}{\displaystyle}

\newcommand{\beq}{\begin{eqnarray}}
\newcommand{\eeq}{\end{eqnarray}}

\newcommand{\Gam}{{\mathit \Gamma}}
\newcommand{\Om}{{\mathit \Omega}}

\renewcommand{\d}{{\rm d}}

\newcommand{\sta}[2]{{\scriptstyle #1 \atop \scriptstyle #2 } }
\newcommand{\Sta}[2]{{ #1 \atop #2 } }

\begin{document}

\title{Light scattering by an oscillating dipole in a focused beam}

\author{Gert Zumofen}
\affiliation{Laboratory of Physical Chemistry, ETH Zurich, 8093 Zurich, 
Switzerland}
\author{Nassiredin M.\ Mojarad}
\affiliation{Laboratory of Physical Chemistry, ETH Zurich, 8093 Zurich, 
Switzerland}
\author{Mario Agio}
\email{mario.agio@phys.chem.ethz.ch}
\affiliation{Laboratory of Physical Chemistry, ETH Zurich, 8093 Zurich, 
Switzerland}


\begin{abstract}
The interaction between a focused beam and a single classical
oscillating dipole or a two-level system located at the focal spot is
investigated.
In particular, the ratio of the scattered to incident power is studied
in terms of the oscillator's scattering cross section and the
effective focal area. Debye diffraction integrals are
applied to calculate it and results are reported for a directional dipolar 
wave. Multipole expansion of the incident beam is then considered and the
equivalence between this and the Debye diffraction approach is discussed.
Finally, the phase change of the electric field upon the interaction with
a single oscillator is studied.
\end{abstract}

\maketitle

\section{Introduction}

The realization of quantum networks and repeaters for quantum information
science crucially depends on an efficient interface between photons
and single quantum systems~\cite{CZH}. Strong interaction has been
achieved by coupling single emitters to optical resonators~\cite{BBM,DPA,SKF}
and has also been predicted for emitters located in waveguides where the
light is tightly confined in the transverse
dimensions~\cite{KC,DHR,SF,CSD,ZGL}. In free space, the question is to
what extent photons may interact with a single oscillating
dipole~\cite{SMK,PI,SAL}.
Recent experiments have demonstrated that light focused on single ions,
molecules, or quantum dots may be attenuated in transmission by a few
percents~\cite{GWB,WGH,GSD,VAD,TCA}.
In a recent theoretical study we have shown that a focused light beam
can be perfectly reflected by a single oscillating dipole located at the
focal spot~\cite{ZNS}.
In this paper we investigate in more detail the cases where the beam is a
focused plane wave and a directional dipole wave. We discuss the equivalence
between the multipole expansion and the Debye diffraction approaches.
Moreover, we compute the phase shift induced on the electric field by the
dipole and find that a few degrees are easily obtainable using realistic
focusing parameters and off-resonance excitation.

The strength of the interaction between a beam and an oscillator can be
expressed by $\cal K$, the ratio of the scattered to incident power.
$\cal K$ can also be given as the ratio of two independent quantities, the
scattering cross section $\sigma$ and the inverse of an effective focal
area $\cal A$~\cite{ZNS,EK}
\beq
  {\cal K}= \frac{P_{\rm sca}}{P_{\rm inc}}
    =   \frac{ \sigma }
    { \cal A  } ~,
\label{eq1}
\eeq
$\sigma$ and $\cal A$ depend exclusively on the oscillator
and focusing setup properties, respectively.
For a classical oscillator and a two-level system (TLS)
the cross section reads
\beq
  \sigma = \left\{ \begin{array}{ll}
    \D  \sigma_0  \frac { \Gam^2 }{ 4 \Delta^2 + \Gam^2}~, &\mbox{classical oscillator} \\[2mm]
  \D \sigma_0  \frac{ \Gam_1^2  }{ 4  \Delta^2 + \Gam_1^2 +  2 \Om^2   }~,
       &\mbox{TLS} ~,
\end{array} \right.
\label{eq2}
\eeq
where we assumed that there is no damping other than by radiation.
$\sigma_0 = 3 \lambda^2/(2 \pi)$ depends solely on the wavelength of the
transition~\cite{Jac}. $\Gam$ results from radiation reaction for
the classical oscillator~\cite{Jac}, while $\Gam_1$ represents the
Einstein coefficient of spontaneous decay~\cite{CDG}. $\Delta$ denotes
the detuning from resonance and $\Om$ is the Rabi frequency of the TLS,
which imposes saturation effects at stronger incident light.

\begin{figure}
\centerline{\includegraphics[width=8cm]{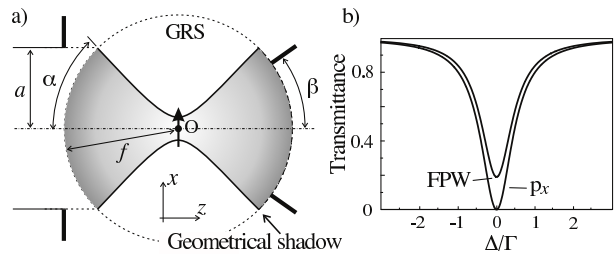}}
\caption{a) The incident light propagating from left to right is
focused onto an oscillating dipole located at the focal spot in vacuo.
GRS: Gaussian reference sphere,
$a$: entrance-aperture radius, $\alpha$: entrance half angle,
$\beta$: collection half angle, $f$: focal length, O: focal spot.
b) Transmittance as a function of the laser detuning is displayed for a
focused plane wave (FPW) with $\alpha=\beta=\pi/3$ and for a directional
dipole wave (p$_x$) with $\alpha=\beta=\pi/2$.}
\end{figure}

For a point-like oscillator the scattered power depends solely on the
field strength at the position of the oscillator. Accounting for the
 electric nature  of the interaction, the effective focal area
can be given as the ratio of the power transmitted through the focal plane (FP)
and the electric energy density at the focal spot~\cite{ZNS}
\beq
  {\cal A} = \frac{ \int_{\rm FP} {S}_z \, \d^2 r }
    { 2 c W_{\rm el}({\rm O}) }
    =  \frac{ \int_{\rm FP} {S}_z \, \d^2 r }{ {S}_z({\rm O})}~,
\label{eq3}
\eeq
where $S_z$ denotes the $z$ component of the Poynting vector in the FP
 and $W_{\rm el}({\rm O})$ is the electric energy density
at the focal spot O~\cite{RW}.
The integration is taken over the FP. The second equality holds
for circular symmetry of the incident field strength with respect
of the  $z$ axis; a condition which, for instance, is obeyed by a
focused plane wave (FPW) but not by a directional dipole wave.
Figure 1a) describes an ideal lens that projects the incident field onto the
Gaussian reference sphere (GRS), which represents the locus of equal
phase of the incoming converging and also for the outgoing diverging
mode. Because the lens is assumed to be in the far-field region, the fields
are tangential on the GRS.

\section{Debye diffraction}

An established approach of calculating the field in the focal area is
provided by the Debye diffraction integrals. This approach was initiated
by Debye using Green's theorem~\cite{Deb} and was extended
by Wolf using the method of stationary phase~\cite{Wol}. For an
incident plane wave the method was extensively applied by Richards and
Wolf~\cite{RW}. These considerations led to the Debye diffraction
integral for the electric and magnetic fields in the focal area~\cite{RW,Sta}
\beq
\label{eq4}
  {\bf E}({\bf r})  =  - \frac{i k}{2 \pi} \int_{\Sigma_{\rm inc}} {\bf A}
      e^{i k {\bf r} \cdot {\bf s} } \d \Sigma ~, \hspace{2cm}
  {\bf H}({\bf r})  =  - \frac{i k}{2 \pi c} \int_{\Sigma_{\rm inc}} {\bf s} \times {\bf A}
      e^{i k {\bf r} \cdot {\bf s} } \d \Sigma ~,
\eeq
where ${\bf A}$ denotes the vectorial angular
spectrum of the incident wave. $k$ is related to the wavelength by
$k=2 \pi/\lambda$. $\bf r$ is the position in the focal area and
${\bf s}$ is the unit vector in the direction of the plane wave.
The integration is carried out over the incident solid angle
$\Sigma_{\rm inc}$ bounded by the semiaperture angle $\alpha$.
Calculations are presented in Fig. 2 for the case of a directional
dipole wave indicated by p$_x$. Such a wave is constructed by considering
the emission pattern at the left hemisphere of the GRS of an electric dipole
located at O and oriented along the $x$ axis and by reversing the
propagation direction~\cite{SD}. The angular spectra of the
FPW~\cite{RW,ST1} and the p$_x$ wave read
\beq
   {\bf A} = \left\{ \begin{array}{ll} f E_0 \sqrt{ \cos \theta}
      \left(  \cos \phi \, {\bf e}_\theta - \sin \phi \, {\bf e}_\phi \right) ~, & \mbox{FPW} \\
      f E_0 \left( \cos \theta \cos \phi \, {\bf e}_\theta - \sin \phi \, {\bf e}_\phi \right) ~, & {\rm p}_x \\
          0  ~, & \theta > \alpha ~.
\end{array} \right.
\label{eq4a}
\eeq
In Figs. 2a) and 2b), the Poynting vector component $S_z$
and the electric energy density proportional to $|E_x|^2$ are displayed
along the $x$ and $y$ axes in the FP. Note that the field components $E_y$
and $E_z$ are zero on these axes. Figure 2c) shows areas of positive and
negative values of $S_z$ in the FP, i.e. areas of forward and
backward propagation. The changes of direction are a signature of field
vortices in the focal area, as reported for a FPW~\cite{Sta,BDW}. In Fig.
2d) the phase of the electric field relative to that of a plane wave is
plotted for positions along the $z$ axis.
One notes a characteristic phase anomaly in the neighborhood of the
focal spot associated with a phase jump of $\pi$, which is also termed
Gouy phase~\cite{BW,HM}. Furthermore there are oscillations, which do not
vanish for increasing $z$ displacements. This behavior is singular
for propagation along the $z$ axis, while for directions
increasingly tilted away from the $z$ axis the oscillations
progressively die out at larger distances~\cite{Deb}.

\begin{figure}
\centerline{\includegraphics[width=9cm]{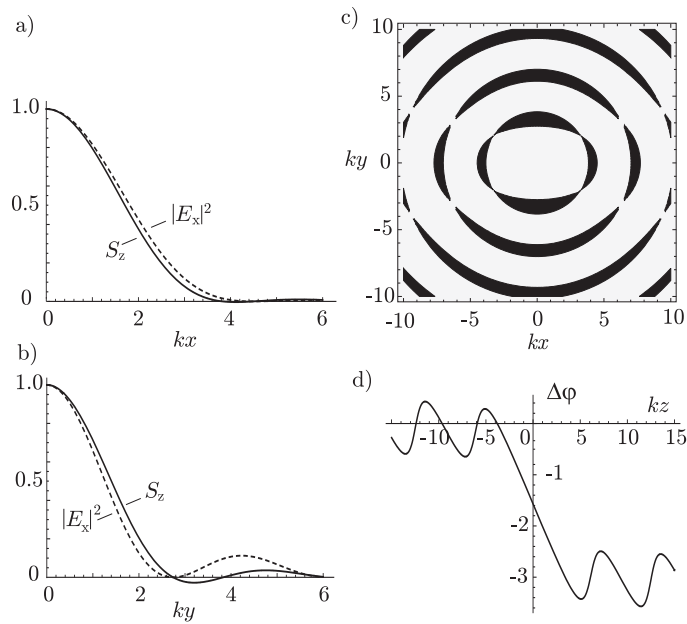}}
\caption{a) The $z$ component of the Poynting vector $S_z$ (full
curve) and the electric energy density given as $|E_x|^2$
(dashed curve) of a p$_x$ wave along the $x$ axis in the FP
and normalized to their respective values at $x=0$. b) Same as a)
but along the $y$ axis.
c) Contour plot of $S_z$ in the FP. Bright and black areas refer
to the Poynting vector in the positive and negative $z$ direction,
respectively.
d) Phase of the focused electric field $E_x$ on the $z$ axis
relative to that of a plane wave. Note the oscillatory behavior that
does not vanish for large $|z|$-values. $\alpha=\pi/2$ in all cases.}
\end{figure}

Using Eq.~(\ref{eq3}) we calculated $\cal A$ for four different cases,
namely for the FPW, the p$_x$ wave, the dipolar wave with the
generating dipole oriented along $z$ axis, and  for combined electric and
magnetic generating dipoles directed along the $x$ and $y$ axis,
respectively~\cite{ZNS}. Here we present, {\it pars pro toto}, the
results for the electric field ${\bf E}({\rm O})$ at the origin, the
effective focal area $\cal A$, and the scattering ratio $\cal K$ for the
p$_x$ wave
\beq
\label{eq5}
 {\bf E}({\rm O}) =  -i  \frac {2 k f E_0}{ 3 }  {\cal F}(\alpha) \,  
{\bf e}_x ~, \hspace{1cm}
  \frac 1 {\cal A} =    \frac {k^2}{ 3 \pi} {\cal F}(\alpha)  ~, \hspace{1cm}
  {\cal K}_0 &=&  2  {\cal F}(\alpha)  ~,
\eeq
where  the subscript of ${\cal K}_0$ indicates
that $\sigma = \sigma_0 $ is assumed. The p$_x$ wave has the property
that the three quantities depend in the same way on the semiaperture
angle $\alpha$ through
\beq
 {\cal F}(\alpha) = \frac 14 \left( 4 - 3 \cos \alpha - \cos^3 \alpha \right) ~.
\eeq
As pointed out in Ref.~\cite{ZNS}, ${\cal K}_0$ reaches for
$\alpha=\pi/2$ the maximum possible value of 2, which also establishes
the maximum possible scattering ratio for a directional focused beam in free
space. ${\cal K}_0>1$ indicates that the scattered power is larger than
the incident power. However, this does not violate the energy conservation law
because of destructive interference in the forward direction.
Taking the interference into account the transmittance
$\cal T$, the ratio of the transmitted and incident power, is given by
\beq
   {\cal T} = 1 - {\cal R} = 1 - \frac 12 \frac {\sigma}{ \cal A}~,
\label{eq6}
\eeq
where $\cal R$ is the reflectance, the ratio of
the back scattered to incident power. The factor of 1/2 in the
second equality accounts for the fact that equal amount of
scattering takes place in the forward and backward directions.
Based on the procedure outlined in Ref.~\cite{ZNS} we also
determined the transmittance as a function of the
semiaperture angle $\alpha$ and semicollection angle $\beta$
\beq
\label{eq7}
{\cal T}_0(\alpha,\beta) =  1 - \frac1{16}
\left(4 - 3 \cos\alpha - \cos^3 \alpha \right) 
     \left(4 + 3 \cos\left(\max\{\alpha, \beta\}\right)
        + \cos^3\left( \max\{\alpha,\beta\} \right)\right) ~,
\eeq
where the subscript to ${\cal T}_0$ indicates that $\sigma=\sigma_0$ is
assumed. Examples of $\cal T$ as a function of the detuning are presented
in Fig. 1b) for the FPW and for the p$_x$ wave. Figure~3 displays
a rapid decrease of ${\cal T}_0$ with increasing
$\alpha$ and an edge along the geometrical shadow boundary $\alpha=\beta$,
as for the FPW~\cite{ZNS}. As shown in Eq.~(\ref{eq7}), ${\cal T}_0$ is
invariant with respect to $\beta$ for $\beta<\alpha$, while for
$\beta>\alpha$, ${\cal T}_0$ increases with $\beta$.
Contrary to the FPW, ${\cal T}_0(\alpha,\beta)$ decreases
monotonously with increasing $\alpha$ and reaches the value of zero at
$\alpha=\pi/2$.

\begin{figure}
\centerline{\includegraphics[width=7cm]{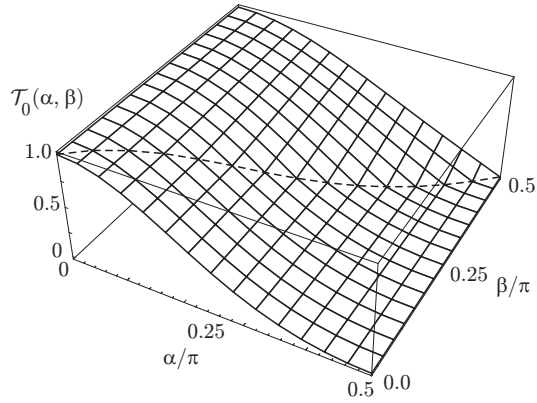}} \caption{
Transmittance $ {\cal T}_0$ of a p$_x$ wave as a
function of the angles $\alpha$ and $\beta$ as defined in Fig.~1a.
The dashed curve indicates the edge along the geometrical shadow
boundary $\alpha=\beta$. }
\end{figure}

\section{Multipole expansion}

Another approach convenient for the description of focused fields
is given by a multipole expansion~\cite{ST1,Str,BH,vEn,MSA}. Adopting
the notation of Bohren and Huffman~\cite{BH} we write for the
electric field in the most general form
\beq
\label{eq8}
  {\bf E}({\bf r}) = \sum\limits_{\ell}
   \sum\limits_{m=0}^\ell & & \left( B_{e,m,\ell} {\bf M}_{e,m,\ell}({\bf r})
     +  A_{e,m,\ell} {\bf N}_{e,m,\ell}({\bf r})  + e \rightarrow o \right)~,
\eeq
where ${\bf M}_{\sta{e}{o},m,\ell}({\bf r})$ and
${\bf N}_{\sta{e}{o},m,\ell}({\bf r})$ are real valued and denote
complete sets of magnetic and electric multipoles. $B_{\sta{e}{o},m,\ell}$
and $A_{\sta{e}{o},m,\ell}$ are the corresponding coefficients. Assuming a
linearly polarized field in front of the incident lens and
aligning the $x$ axes to the incident field polarization
the expansion in Eq.~(\ref{eq8}) can be restricted to
${\bf M}_{o,1,\ell}({\bf r})$ and ${\bf N}_{e,1,\ell}({\bf r})$
multipoles for the electric field and to ${\bf M}_{e,1,\ell}({\bf r})$
and ${\bf N}_{o,1,\ell}({\bf r})$ for the magnetic field, respectively.

The calculation of the coefficients requires some
attention~\cite{NRH}. The source-free field mode may be considered as a
sum of the converging incoming and
diverging outgoing field. Because the outgoing mode is purely a consequence of
the incoming mode, only the latter is needed for a unique
determination of the coefficients. This concept was applied
for instance by Sheppard and T\"or\"ok~\cite{ST1},
where the multipoles of the expansion were associated with spherical
Hankel functions.
The direct expansion in terms of multipoles for the source-free field is
also possible. However, in this case the converging field at the entrance
and the diverging field at the exit of the GRS have to be taken into account.
For this purpose the field symmetry on the GRS has to be
considered~\cite{Wol1,CW}, which can be derived from the Debye scattering
integrals in Eqs.~(\ref{eq4}) when assuming positions 
diametral with respect to the origin
\beq
   {\bf E}(- {\bf r}) & = & - \frac{ik}{2 \pi} \int_{\Sigma_{\rm inc}} {\bf A}
     e^{-ik {\bf r \cdot s}} \, \d \Sigma 
     = \left[  \frac{ik}{2 \pi} \int_{\Sigma_{\rm inc}} {\bf A}
     e^{ik {\bf r \cdot s}} \, \d \Sigma  \right]^* = - {\bf E}^*( {\bf r})~,
\eeq
which means that the field is antihermitian for diametral positions on the
GRS. The corresponding relationship of the field's phase $\varphi$ reads
\beq
   \varphi(x,y,z) = - \varphi(-x,-y,-z) - \pi,~{\rm mod}~  2 \pi~.
\eeq
The phase shift of $-\pi$ demonstrates the phase anomaly in the neighborhood
of the focal spot (see Fig. 2d) and it is equal to the Gouy phase
acquired when the beam traverses the focus.

\begin{figure}
\centerline{\includegraphics[width=4.5cm]{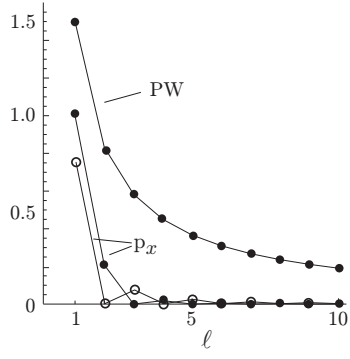}}
\caption{Multipole expansion coefficients for the non-focused plane wave
(PW) and the p$_x$ wave. $|A_{e,1,\ell}|/E_0 = |B_{o,1,\ell}|/E_0 $
(dots) are presented for the PW and $|A_{e,1,\ell}|/(E_0 f k)$ (dots) and
$|B_{e,1,\ell}|/(E_0 f k)$ (circles) for the p$_x$ wave, respectively.
For the latter a semiaperture angle of $\alpha=\pi/2$ was assumed.}
\end{figure}

Here we follow the approach of Borghi~\cite{Bor} and Borghi {\it et
al.}~\cite{BSA} and expand the angular spectrum $\bf A$ of the incident
field in surface vector harmonics and substitute the expansion into Eqs.~(4).
This procedure assures that the Debye-diffraction and multipole-expansion
method are literally the same. We write for the expansion
\beq
  {\bf A} = \frac{(-i)^\ell}{2 k}  \sum\limits_{\ell=1}^\infty 
\left( B_{o,1,\ell}  {\bf \widetilde M}_{e,1,\ell}
     + i A_{e,1,\ell}  {\bf \widetilde N}_{e,1,\ell}  \right)~,
\label{eq9}
\eeq
where  ${\bf \widetilde M}_{o,m,\ell}(\theta, \phi)$ and ${\bf \widetilde
N}_{e,m,\ell}(\theta, \phi)$ are real valued and complete sets of vectorial
surface harmonics, which are independent on the radial variable $r$.
They are related to the spherical vector harmonics ${\bf Y}$ and $\bf Z$
by~\cite{Jac,LegendreP}
\beq \begin{array}{lll}
   {\bf Y}_\ell^{ 1} &=& \D i \left(\frac{2 \ell +1 } { 2 \pi \ell(\ell+1)} \right)^{1/2}
     \left(  {\bf \widetilde M}_{e,1,\ell} +  i \, {\bf \widetilde M}_{o,1,\ell} \right)  ~, \\[2mm]
   {\bf Z}_\ell^{1} &=& \D i \left(\frac{2 \ell +1 } { 2 \pi \ell(\ell+1)} \right)^{1/2}
    \left(  {\bf \widetilde N}_{e,1,\ell} +  i \,  {\bf\widetilde  N}_{o,1,\ell} \right)~,
\end{array}
\label{eq10}
\eeq
where ${\bf Z}_\ell^{ 1} = {\bf s} \times  {\bf Y}_\ell^{1}$.
Because of the completeness and orthogonality of the basis functions,
the coefficients are given by
\beq
\begin{array}{lll}
   B_{\sta{e}{o},1,\ell} &= \D 2 k  i^{\ell -1} \frac{ 2 \ell +1}{ 2 \pi  \ell^2 (\ell+1)^2} \int_{\Sigma_{\rm inc}} {\bf A}
      \cdot {\bf \widetilde M}_{\sta{e}{o},1,\ell} \, \d \Sigma ~,\\[3mm] \nonumber
   A_{\sta{e}{o},1,\ell} &= \D- 2 k  i^{\ell} \frac{ 2 \ell +1}{ 2 \pi  \ell^2 (\ell +1)^2}
         \int_{\Sigma_{\rm inc}} {\bf A} \cdot {\bf \widetilde N}_{\sta{e}{o},1,\ell}
         \, \d \Sigma ~,
\end{array}
\label{e11}
\eeq
where the prefactors result from normalization of the basis functions and
from accounting of the Whittaker type of transformation, which provides
a relationship between surface vector harmonics and multipoles~\cite{Whi,DW}.
For the magnetic multipoles this relationship reads
\beq
    {\bf M}_{\sta{e}{o},1,\ell}({\bf r}) & = & \frac{(-i)^\ell}{4 \pi}
       \int_{4 \pi} {\bf \widetilde M}_{\sta{e}{o},1,\ell}({\bf s})
        e^{i k {\bf s}\cdot {\bf r}} \d \Sigma
 = j_\ell(k r) {\bf \widetilde M}_{\sta{e}{o},1,\ell}(\theta,\phi)~,
\label{eq12}
\eeq
and analogously for the electric multipoles
\beq
  {\bf N}_{\sta{e}{o},1,\ell}({\bf r})&=&
\frac 1 k \nabla \times {\bf M}_{\sta{e}{o},1,\ell}({\bf r})
  =  \frac{(-i)^{\ell-1}}{4 \pi}  \int_{4 \pi} {\bf s} \times {\bf 
\widetilde M}_{\sta{e}{o},1,\ell}({\bf s})
        e^{i k {\bf s}\cdot {\bf r}} \d \Sigma
  = \frac{(-i)^{\ell-1}}{4 \pi}  \int_{4 \pi} {\bf \widetilde N}_{\sta{e}{o},1,\ell}({\bf s})
        e^{i k {\bf s}\cdot {\bf r}} \d \Sigma ~.
\label{eq14}
\eeq
We finally write the fields in terms of multipoles
\begin{eqnarray}
{\bf E}({\bf r}) &  = & \sum \limits_{\ell=1}^\infty  \left( B_{o,1,\ell}  
{\bf M}_{o,1,\ell}({\bf r})
+ A_{e,1,\ell}  {\bf N}_{e,1,\ell}({\bf r})  \right)  ~, \\
{\bf H}({\bf r}) & = & \frac {-i}c \sum \limits_{\ell=1}^\infty  \left( 
B_{o,1,\ell}  {\bf N}_{o,1,\ell}({\bf r})
+  A_{e,1,\ell}  {\bf M}_{e,1,\ell}({\bf r})   \right) ~,
\label{eq15}
\end{eqnarray}
and for completeness we also present expressions for the surface vector
harmonics and multipoles
\begin{eqnarray}
\nonumber
    \widetilde {\bf M}_{\sta{e}{o}, 1,\ell}  &=& 
\D \pi_\ell  \Sta{-\sin \phi}{~\cos \phi}  \, {\bf e}_\theta
       -  \tau_\ell  \Sta{\cos \phi}{\sin \phi}  \, {\bf e}_\phi   ~, \\[4mm] \nonumber
     \widetilde {\bf N}_{\sta{e}{o}, 1,\ell} &=&  
\D \tau_\ell  \Sta{\cos \phi}{ \sin \phi}  \, {\bf e}_\theta
      +  \pi_\ell  \Sta{- \sin \phi}{~\cos \phi}  \, {\bf e}_\phi ~, 
\\[4mm] \nonumber
{\bf M}_{\sta{e}{o}, 1,\ell}  &=&
\D j_\ell  \widetilde {\bf M}_{\sta{e}{o}, 1,\ell} ~, \\[4mm] \nonumber 
{\bf N}_{\sta{e}{o}, 1,\ell}  &=& 
\D \frac 1{kr}   \left( \ell(\ell+1) j_\ell \pi_\ell  \Sta{ \cos \phi}{ \sin \phi}   \, {\bf e}_r
     + S_\ell \, \widetilde {\bf N}_{\sta{e}{o}, 1,\ell}   \right) ~,
\end{eqnarray}
where $j_\ell= j_\ell(kr)$ is the spherical Bessel function. $S_\ell$
is a related to $j_\ell$, and $\pi_\ell$ and $\tau_\ell$ follow from the
Legendre polynomials $P_\ell^1$ as
\beq
  S_\ell =  \frac{\d  ( kr  j_\ell (kr)) }{ \d (kr) } ~, \hspace{1cm}
  \pi_\ell = \frac{ P_\ell^1( \cos \theta) }{ \sin \theta} ~, \hspace{1cm}
    \tau_\ell = \frac{ \d P_\ell^1( \cos \theta) }{ \d  \theta} ~.
\eeq

In Fig. 4 we depict the coefficients $A_{e,1,\ell}$ and
$B_{o,1,\ell}$ of the p$_x$ wave for
$\alpha=\pi/2$, and compare them with the coefficients of a
non-focused plane wave (PW)~\cite{BH}. We note that $A_{e,1,\ell}$ differs
from zero for even $\ell$ except for $\ell=1$ while the
$B_{o,1,\ell}$ coefficients differ from zero exclusively for
odd $\ell$. The fact that coefficients with $\ell>1$ do not vanish
for the p$_x$ wave is somewhat surprising. However, these are required
to maintain the propagation characteristics of a directional wave and
to guarantee power conservation throughout the space on the basis
of a source-free focused field. Figure 5 displays the quality of
the expansion for the FPW and p$_x$ wave when the
number of terms is truncated. It is apparent that quite a few
terms are required for a decent reproduction of the angular
spectra for $\alpha=\pi/2$ and even more terms are required for
$\alpha < \pi/2$.

\begin{figure}
\centerline{\includegraphics[width=12cm]{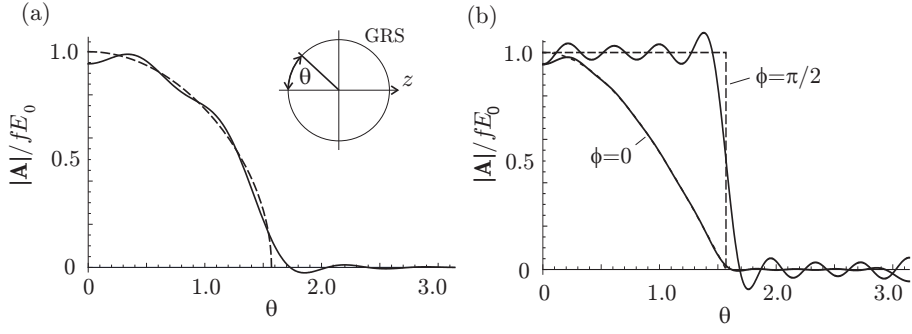}}
\caption{Angular spectra $\bf A$ of the incident field according to
Eq.~(\ref{eq4a}) and approximations according to Eq.~(\ref{eq9}) are
given by dashed and full lines, respectively, for $\alpha=\pi/2$.
(a) FPW. (b) Dependence along the $x$ and $y$ direction, indicated by
$\phi = 0$ and $\phi = \pi/2$, respectively, for the p$_x$ wave.
The inset indicates the angle $\theta$ as it is considered for this figure.
The summation of terms is truncated to $\ell_{\rm max} = 8$ and 16 in
(a) and (b), respectively.}
\end{figure}

Of special interest is the property that all multipoles are zero at the
origin  except the electric dipole mode ${\bf N}_{e,1,1}^{(1)}$.
Therefore, in cases where the field at the origin  only is relevant, the
analysis can be simplified by decomposing the incident field into dipolar
and nondipolar modes. This concept was introduced by van Enk~\cite{vEn}
and was considered in Refs.~\cite{PI,ZNS}. ${\bf N}_{e,1,1}$ at the
origin and in the far-field region reads~\cite{BH}
\beq
\label{eq18}
\begin{array}{lll}
{\bf N}_{e,1,1}^{(1)} & = \frac 23 {\bf \hat e}_x , & {\bf r} = {\rm O} 
~, \\[3mm] \D {\bf N}_{e,1,1}^{(3)} & = \frac{ \D e^{ i (kr-\pi/2)}}{kr} 
\left( \cos \theta \cos \phi \, {\bf \hat e}_\theta - \sin \phi \, {\bf \hat e}_\phi \right) , 
& kr \gg 1 ,~z>0 ~,
\end{array}
\eeq
where the superscripts (1) and (3) denote the
source-free field with the spherical Bessel function and the outgoing
mode with the spherical Hankel function, respectively. The concept of field
decomposition into dipolar and nondipolar components becomes
particularly useful when the scattered field ${\bf E}_{\rm sca}$
is considered, which in the case of a classical dipole oriented along
the $x$ axis is given by
\beq
\label{eq19}
   {\bf E}_{\rm sca}({\bf r}) = -\frac 32 E_{\rm inc}({\rm O}) 
\frac{\Gam}{2 \Delta + i \Gam} \frac{e^{i k r} }{k r}
\left( \cos \theta \cos \phi \, {\bf \hat e}_\theta
   - \sin \phi \, {\bf \hat e}_\phi \right) , \hspace{1cm} kr \gg 1 ~.
\eeq
Inserting the expression
\beq
   E_{\rm inc}({\rm O}) = A_{ e,1,1} {\bf N}_{e,1,1}^{(1)}({\rm O}) \cdot {\bf e}_{x}~,
\label{eq20}
\eeq
into Eq.~(\ref{eq19}), it is fairly easy to see that at resonance the
dipole component of the incident field is exactly canceled by the
scattered field in the forward direction. Therefore, the outgoing field is
given by
\beq
    {\bf E}_{\rm out}  =  {\bf E}_{\rm inc} + {\bf E}_{\rm sca}
     =  {\bf E}_{\rm inc} -  A_{ e,1,1}  \frac{\Gam}{2 \Delta + i \Gam}
    \frac{e^{i k r} }{k r} \left( \cos \theta \cos \phi \, {\bf \hat e}_\theta
   - \sin \phi \, {\bf \hat e}_\phi \right)~,
\label{eq21}
\eeq
where, apart from ${\bf E}_{\rm inc}$, the coefficient $A_{e,1,1}$ is
the only variable that depends on the
focusing specifications.
For the FPW and the p$_x$ wave,  $A_{ e,1,1} $ is given analytically as
\beq
    A_{ e,1,1}
     = \left\{ \begin{array}{ll}  \D -i \frac{1 }{10} f k E_0
      \left( 8 - \cos^{3/2}\alpha \left( 5 + 3 \cos \alpha\right)\right)
        ~, &\mbox{FPW} \\ [3mm]
     \D  -i \frac 1{4} f k E_0
      \left( 4 - 3 \cos \alpha -  \cos^3 \alpha \right)
      ,&  \mbox{ p}_{x} ~.
       \end{array} \right.
\label{eq22}
\eeq
For a p$_x$ wave with $\alpha=\pi/2$, $A_{ e,1,1}/( f k E_0) =1$ as shown in
Fig. 4. Furthermore, by inserting Eqs.~({\ref{eq18}) and
({\ref{eq22}) into Eq.~(\ref{eq20}) one sees that the electric field at the origin is
the same as in Eq.~(\ref{eq5}), confirming that the
Debye-diffraction and the multipole-expansion methods yield identical
results.

\begin{figure}
\centerline{\includegraphics[width=5cm]{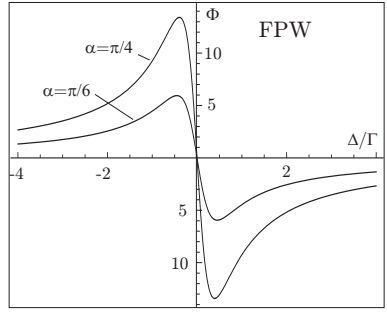}}
\caption{Phase shift $\Phi$ of Eq.~(\ref{eq24}) in units of degrees
as a function of the laser detuning. Results are plotted for the FPW and
for two different semiaperture angles $\alpha$, as indicated.}
\end{figure}

\section{Phase shift by scattering}

With the above considerations it is also easy to calculate the phase shift
$\Phi$ imposed on the beam by a single oscillator at the focal spot. The
phase shift $\Phi$ is defined by
\beq
   \Phi = {\rm arg}\left( {\bf E}_{\rm out} \cdot {\bf E}_{\rm inc}^*\right)~,
\label{eq23}
\eeq
where ${\bf E}_{\rm inc}^*$ is introduced as a reference field for the
detection of the phase shift. Making use of Eqs.~(\ref{eq21}) and
(\ref{eq22}) and assuming that a detector is positioned on the $z$ axis,
we find
\beq
 \Phi =  {\rm arg}\left(1 - \frac{i \Gam }{2 \Delta + i \Gam} ~\times
         \begin{array}{ll}    \frac{1 }{10}
      \left( 8 - \cos^{3/2}\alpha \left( 5 + 3 \cos \alpha\right)\right)
        \\ [2mm]
       \frac 14\left( 4 - 3 \cos \alpha -  \cos^3 \alpha \right)
       \end{array} \right)  \begin{array}{ll} ,\mbox{ FPW}   \\ [2mm] ,\mbox{ p}_{x}~, \end{array}
\label{eq24}
\eeq
where an extra negative sign is introduced to account for the
Gouy phase shift of $\pi$ imposed on the incident field.
$\Phi$ in Fig.~6 shows a typical dispersive of behavior. It amounts
to 5-15 degrees at the extremal points for semiaperture angles accessible
in experiments. The extremal points are located at approximately
$\Delta/\Gam = 1/2$ and $\Phi$ decays only slowly with increasing
detuning. We expect that integration over a collection solid angle would
not change substantially the picture gained from Eq.~(\ref{eq24})
because of the coinciding phase fronts of the incident and scattered field.

\section{Conclusions}

We studied the scattering of a FPW and a p$_x$ wave
by a single oscillator, with emphasis on the equivalence between the
Debye diffraction and multipole expansion approaches. We
systematically applied the concept of the GRS as the locus of
equal phases in the forward and backward direction and paid special
attention to the calculation of the multipole expansion
coefficients on the basis of source-free fields. We thus derived
an analytical expression ${\cal T}_0(\alpha,\beta)$ for the transmittance
of a p$_x$ wave.
We finally demonstrated that a considerable phase shift of a few degrees
is imposed on the light beam by a single oscillator at a detuning
significantly larger than the linewidth. This property, for instance,
might be exploited for the non-resonant detection of single emitters.

\section*{Acknowledgments}

We thank V. Sandoghdar for fruitful discussions and encouragement.
This work was supported by the Swiss National Science Foundation and by the
ETH Zurich research grant TH-49/06-1.

\end{document}